# Magnetic anisotropy reversal driven by structural symmetry-breaking in monolayer α-RuCl$_3$

Bowen Yang[1,2], Yin Min Goh[3], Suk Hyun Sung[4], Gaihua Ye[5], Sananda Biswas[6], David A. S. Kaib[6], Ramesh Dhakal[7], Shaohua Yan[8], Chenghe Li[8], Shengwei Jiang[9,10], Fangchu Chen[1,2], Hechang Lei[8], Rui He[5], Roser Valentí[6], Stephen M. Winter[7*], Robert Hovden[4,11*], Adam W. Tsen[1,12*]

[1]Institute for Quantum Computing, University of Waterloo, Waterloo, ON N2L 3G1, Canada.

[2]Department of Physics and Astronomy, University of Waterloo, Waterloo, ON N2L 3G1, Canada.

[3]Department of Physics, University of Michigan, Ann Arbor, MI 48109, USA.

[4]Department of Materials Science and Engineering, University of Michigan, Ann Arbor, MI 48109, USA.

[5]Department of Electrical and Computer Engineering, Texas Tech University, Lubbock, TX 79409, USA.

[6]Institut für Theoretische Physik, Goethe-Universität Frankfurt, Max-von-Laue-Strasse 1, 60438 Frankfurt am Main, Germany

[7]Department of Physics and Center for Functional Materials, Wake Forest University, Winston-Salem, NC 27109, USA

[8]Department of Physics and Beijing Key Laboratory of Opto-electronic Functional Materials & Micro-nano Devices, Renmin University of China, 100872 Beijing, China.

[9]Department of Physics, Cornell University, Ithaca, NY, USA.

[10]Key Laboratory of Artificial Structures and Quantum Control (Ministry of Education), School of Physics and Astronomy, Shanghai Jiao Tong University, Shanghai, China

[11]Applied Physics Program, University of Michigan, Ann Arbor, MI 48109, USA.

[12]Department of Chemistry, University of Waterloo, Waterloo, ON N2L 3G1, Canada.

*Correspondence to: awtsen@uwaterloo.ca, hovden@umich.edu, winters@wfu.edu

The Kitaev model is a celebrated spin-½ model on the two-dimensional (2D) honeycomb lattice with bond-dependent Ising interactions[1], which features a highly entangled quantum spin liquid (QSL) ground state, fractionalized Majorana excitations, and a series of magnetic-field-induced quantum phase transitions[1–7]. The search for materials realizing the Kitaev model has been an ongoing challenge for over a decade and may potentially lead to applications in fault-tolerant topological quantum computing[8]. Yet, the unavoidable presence of non-Kitaev interactions (Heisenberg, off-diagonal, next-nearest neighbor, etc.) almost always drives the ground state away from the QSL phase, and a careful tuning of the exchange parameters is needed[9–12]. α-RuCl$_3$ is a layered van der Waals material that is a particularly promising candidate to realize Kitaev physics[13–15]. Although the ground state is zigzag (ZZ) antiferromagnetic (AFM), this ordering can be suppressed with the application of an ~ 6-8T in-plane magnetic field. The presence of a half-integer thermal quantum Hall effect has been reported in this intermediate phase at low temperature[16,17], while an unusual continuum of magnetic excitations can be seen even without magnetic field that persists far above the Néel temperature ($T_N$ ~ 7-8K)[18,19]. Both observations hint at α-RuCl$_3$ being in proximity to a QSL, making it a current subject of intense scrutiny. Yet from a theoretical point of view, a QSL induced by in-plane field generally cannot be accounted for, as most calculations for α-RuCl$_3$ show Kitaev phases more broadly emerging from an out-of-plane field[6,20–23]. Due to the strong "easy-plane" magnetic anisotropy of bulk crystals, however, prohibitively high fields above 30T are required to access such states[24–26].

The strong coupling between the spin, charge, and lattice degrees of freedom in α-RuCl$_3$ presents an exciting opportunity to tune its magnetic interactions via external perturbations. Although pressure[27–29], strain[30], and doping[31] have all been previously used and/or suggested to manipulate the magnetic order in α-RuCl$_3$, the role of dimensionality and interlayer coupling have not been carefully examined. Moreover, pure monolayer systems are in principle expected to more closely realize the Kitaev model compared with their bulk counterparts[32]. In this work, we systematically measure the magnetic excitations in monolayer (1L), bilayer (2L), and trilayer (3L) samples using inelastic electron tunneling spectroscopy (IETS)[33,34]. While the ZZ state persists down to 1L and the magnetic continuum can be seen in a 3L, we observe a clear change in the magnetic anisotropy from easy-plane ($B_\parallel^c \sim 6T$) to easy-axis ($B_\perp^c \sim 6.5T$) with reduced thickness. This change is confirmed by lateral transport measurements upon doping with electrostatic gates. By combining three-dimensional (3D) electron diffraction measurements with *ab initio* calculations, we attribute this effect predominantly to an in-plane distortion of the Cl atoms in monolayer form. We analyze the microscopic exchange parameters for the experimentally determined structure of 1L α-RuCl$_3$ and find that the sign of the average off-diagonal (non-Kitaev) exchange terms, indicative of the magnetic easy axis, is reversed relative to the bulk crystal. This both places the monolayer closer to a regime with highly competing magnetic ground states, the intersection of which may potentially give rise to a QSL, and allows field-tuned transitions in the theoretically compatible out-of-plane direction to be more within reach experimentally. Our work provides a new avenue to tune the magnetic interactions in α-RuCl$_3$ and opens the door to the possible exploration of Kitaev physics in the true 2D limit.

We start by exfoliating α-RuCl$_3$ crystals on oxidized silicon wafers within a nitrogen-filled glovebox and identifying their thickness by optical reflection contrast. To confirm that the thinnest flakes are indeed monolayers, we pick up these samples, encapsulate them with monolayer graphene, and transfer them to 10-nm-thick silicon nitride membranes for 3D electron diffraction measurements (see Methods). Figure 1a shows an electron diffraction pattern of such a structure. Some of the fundamental Bragg peaks of α-RuCl$_3$ used for determining the monolayer structure are circled, although the graphene peaks (along the thick gray circle) can be seen as well. Measuring relative to the graphene peaks, the in-plane lattice constant of our exfoliated α-RuCl$_3$ is determined to be 5.9981 – 6.0088Å, which is consistent with the value for the bulk crystal and thus indicates negligible overall strain[14,35,36]. By tilting the sample, we can measure the diffraction spots as a function of out-of-plane crystal momentum ($k_z$). A sideview schematic of the Bragg rod structure for several of the 1L α-RuCl$_3$ peaks is shown in Fig. 1b and the experimental Bragg rod intensities are shown in Fig. 1c as discrete points together with their expected values in solid lines. In particular, the ($1\bar{2}10$) and ($\bar{1}2\bar{1}0$) peaks exhibit a reduction of symmetry from the ideal crystal. As the $k_z$ dependence for 2L and 3L crystals are markedly different (see Supplementary Section 1), we can confirm our ability to exfoliate and encapsulate α-RuCl$_3$ crystals down to monolayer thickness.

It has been previously demonstrated that IETS is a powerful tool to probe for spin waves in ultrathin insulating magnets in the ~1-10meV range[37–39], the same energy window where various magnetic excitations have been observed in bulk α-RuCl$_3$[18,19,36,40–52]. We thus fabricate a series of metal/α-RuCl$_3$/metal tunnel junctions in inert atmosphere to carry out temperature- and magnetic-field-dependent IETS on 1L, 2L, and 3L α-RuCl$_3$ samples (see Methods). To maximize inelastic electron tunneling, the metal should possess a sizeable Fermi surface with substantial density of states[53]. We mostly use ultrathin (< 10nm) T$_d$-MoTe$_2$ as our metal electrode, although graphene shows qualitatively similar behavior (see Supplementary Section 2). A sideview illustration of our device and measurement geometry is shown in Fig. 1d and a colorized optical image of a representative device is shown in Fig. 1e. Hexagonal boron nitride (h-BN) flakes are used as encapsulation layers for protection.

The upper panel of Fig. 1f shows the measured AC conductance (dI/dV) of a 1L tunnel junction at 2K as a function of the DC voltage. Subtle steps in the curve can be seen centered at ~±1mV, which can be interpreted as increases in the tunneling conductance when the potential difference across the electrodes reaches the energy of a particular inelastic excitation in α-RuCl$_3$[33,34,37,39]. These can be seen more clearly as peaks in the numerical derivative (d$^2$I/dV$^2$) shown in the lower panel of the Fig. 1f. To extract the position and shape of the peaks, we fit them to a pair of Lorentzians (blue) on top of a background (gray)[54]. The resultant fitting is shown in red, which closely traces the experimental result. We next investigate the dependence of these excitation peaks on temperature across different thickness samples to understand whether they are of magnetic or phononic nature.

Figure 2a shows the normalized and background-subtracted d$^2$I/dV$^2$ spectrum for 1L, 2L, and 3L devices from 2K to 10K in a 2D false-color plot for positive bias. The trace at base temperature is overlaid in blue as a reference. The mode at ~1meV appears in all three devices at low temperature and disappears above ~8K. This is near $T_N$ (7K to 8K) measured for bulk crystals of high quality without stacking faults[36,55,56], which suggests a magnetic origin. Spin wave calculations based on *ab initio* studies show the low-energy single magnon in the ZZ AFM phase to be near 1meV at the Y and M points[57], while several experiments have reported bulk magnons near this energy[43,46,47,51]. The energy of our observed mode is thus further consistent with that of an excitation of the ZZ AFM order. The smaller overall conductance of the thicker 3L device allows us to probe IETS to higher voltages. Between ~5-10meV, a broad excitation spectrum is observed that persists up to the highest temperature measured with no apparent discontinuity at $T_N$. This is consistent with the continuum excitations identified in bulk crystals by Raman and neutron scattering, which have been discussed to be connected to fractionalized and/or incoherent excitations[18,19,43,44]. Our results thus show that such unconventional magnetic signatures persist down to at least 3L samples.

To determine $T_N$ more precisely for different thicknesses, we start by fitting Lorentzians to the low-energy mode in the manner described above. This function is known to be a convolution of the intrinsic spectral weight with a temperature-dependent thermal broadening function, $\chi(V) = \frac{1}{kT}\exp(x)\frac{(x-2)\exp(x)+x+2}{(\exp(x)-1)^3}$, where $x = eV/kT$, and a temperature-independent instrument broadening function[33]. The latter is negligible for our measurement conditions (see Supplementary Section 3). We thus extract the intrinsic peak by deconvolving the fitted experimental curve with $\chi$ and integrating the resultant intensity. This value is plotted as a function of temperature in Fig. 2b for the 3L sample. The intercept of a linear fit applied to the data at low temperature yields $T_N \sim 7K$. We apply the same procedure to the other devices to extract $T_N$ as a function of sample thickness in Fig. 2c. The range of $T_N$ measured for high-quality bulk crystals is marked by the gray band. Unlike Heisenberg(-like) magnets obeying the Mermin-Wagner theorem[58], the critical temperature for α-RuCl$_3$ remains essentially unchanged down to monolayer.

In bulk α-RuCl$_3$, magnons can evolve nonmonotonically with the application of an in-plane magnetic field[41,44,49,51]. For example, the magnons at the Γ point first shift down to lower energies with increasing field, reaching a minimum at ~6-8T before shifting up. This critical field has been suggested to host an intermediate QSL region (between the ZZ ground state and high-field paramagnetic state)[16,17], which remains controversial, in part because theoretical studies have only identified models with QSL phases induced by out-of-plane fields[6,20–23]. Due to the easy-plane anisotropy of bulk crystals, however, an out-of-plane field of ≳30T is needed to change the magnetic state, rendering such predicted QSLs largely inaccessible[14,20,24–26,57,59]. We thus proceed to measure the low-energy magnon for all three sample thicknesses with changing magnetic field. In Fig. 3a, we show 1L, 2L, and 3L IETS spectra taken at 2K for $B_\parallel$ between 0 and 14T (in 1T increments) with the traces offset for clarity. The thin gray lines are guides-to-eye for the magnon evolution. To determine the magnon energies more quantitatively, we have performed a Lorentzian fit for each trace and the extracted peak positions are shown in Fig. 3b with changing $B_\parallel$. The 2L and 3L samples show qualitatively similar characteristics—with increasing field, the magnon energy first decreases and then increases, although the field at which the minimum energy occurs appears to be slightly larger for 2L. In contrast, the magnon for 1L is essentially unchanged with magnetic field up to 14T, which suggests that the critical in-plane field necessary to drive the monolayer out of the ZZ state is pushed to a substantially higher value. This trend is captured by the thick blue line. We further note that the observed magnon stiffening for 1L appears to be independent of whether the field is directed along either of the two in-plane crystalline axes (see Supplementary Section 4).

Figure 3c shows the out-of-plane field dependence of the IETS spectra. Here, an opposite trend is observed with changing thickness. The 3L has the stiffest response, consistent with the result for bulk crystals[40], while the low-energy peak position for both 2L and 1L exhibit more curvature with field. Interestingly, a secondary peak at higher energy also develops for the latter samples at finite fields (see guides-to-eye in purple). We have fit all the observed peaks to Lorentzians, and the positions are plotted in Fig. 3d as a function of $B_\perp$. At high fields, the secondary peaks appear to shift with field at roughly twice the rate compared with the low-energy magnon, suggesting that they may originate from two-magnon excitation[44,45,57]. The larger curvature exhibited by this higher energy mode also allows us to clearly identify the critical field for which the energy is minimum—it shifts to higher values with increasing thickness. This trend is captured by the thick orange line and is consistent with the extremely large out-of-plane critical field expected for the bulk crystal. Taken together, the results of Fig. 3 suggest that the magnetic anisotropy is reversed from easy-plane for bulk crystals to easy-axis (out-of-plane) for 1L α-RuCl$_3$. Such a change is striking; however, we must ascertain whether it is intrinsic to monolayer samples or a result of proximity to the T$_d$-MoTe$_2$, a system with strong spin-orbit coupling.

To address this issue, we have fabricated an ultrashort two-terminal device for 1L α-RuCl$_3$ with both few-layer graphene electrodes and top and bottom gates to investigate the field dependence of lateral transport. A colorized scanning electron microscope image and sideview schematic of the device are shown in Fig. 4a. The sample is only in contact with hBN across the channel (length ~ 300nm). Figure 4b shows the DC current-voltage dependence at base temperature for different gate values. Due to the insulating nature of α-RuCl$_3$, the sample only shows measurable current at low bias when large positive gate voltages are applied (electron doping). In the most conductive state (V$_{TG}$ = 9V, V$_{BG}$ = 6V), we have measured the AC conductance upon sweeping the magnetic field (both in-plane and out-of-plane) continuously, and the results are plotted in Fig. 4c for several different temperatures. Overall, there is very little change with in-plane field, consistent with this field direction being along the hard axis. In contrast, there is larger change when the field is applied along the easy axis out of plane. Moreover, a marked kink can be seen in the magnetoconductance at B$_\perp$ ~ 6.5T at low temperatures. This coincides with the critical field for the two-magnon feature measured by IETS. Upon raising the temperature, the kink gradually disappears above T$_N$. These results indicate that the magnetic anisotropy reversal in monolayer α-RuCl$_3$ is likely of intrinsic origin as opposed to proximal contact with T$_d$-MoTe$_2$.

It is well-understood that spin moments in α-RuCl$_3$ are strongly coupled to the charge and lattice degrees of freedom[12,30,60–62]. However, as the anisotropy switching is observed in both gated and intrinsic monolayers, a more probable cause is that the structure of 1L α-RuCl$_3$ deviates from that of the bulk crystal. To investigate whether this is the case, we again turn to electron diffraction measurements performed on the monolayer sample. By carefully fitting the k$_z$ dependence for the various Bragg peaks, we observe three primary distortions of the honeycomb lattice of edge-sharing RuCl$_6$ octahedra (see Supplementary Section 1), which are illustrated in Fig. 5a. First, there is an out-of-plane buckling of the Ru atoms, $\Delta\zeta_{Ru}$, discernable from the asymmetric $(01\bar{1}0)$ and $(0\bar{1}10)$ Bragg rods shown in Fig. 5b. Due to negligible overall strain in the lattice (see discussion of Fig. 1a), the in-plane distortion of Ru should be insignificant. Second, there is a change in the c-axis position Cl atoms relative to the Ru atoms, $\lambda_{Cl}$, as well as a third in-plane distortion of the Cl atoms that are opposite for the top and bottom sublayers, $\Delta r_{Cl}$. A table summarizing the experimentally bounded values for these three distortions are shown in Fig. 5a. Some of these distortions have been previously observed on the surfaces of exfoliated α-RuCl$_3$ flakes and have been attributed to vacancies and/or defects despite preparation in inert atmosphere[63]. Here, diffraction provides a precise measure of the average crystal structure and distortions but is much less sensitive to real-space fluctuations.

The reversal of magnetic anisotropy for 1L α-RuCl$_3$ signifies modification of the spin Hamiltonian $\mathcal{H} = \sum_{<i,j>} \mathbf{S}_i \cdot \mathbf{M}_{ij} \cdot \mathbf{S}_j$ due to the observed distortions. For co-planar Ru, the matrix $\mathbf{M}$ (for the z bond) can be expressed as $\begin{pmatrix} J & \Gamma & \Gamma' \\ \Gamma & J & \Gamma' \\ \Gamma' & \Gamma' & J+K \end{pmatrix}$ for nearest-neighbor interactions, where $J$, $K$, and $\Gamma$ ($\Gamma'$) refer to the Heisenberg, Kitaev, and off-diagonal coupling terms, respectively, although a third neighbor Heisenberg term $J_3$ is expected to contribute as well. With Ru buckling, the symmetry of $\mathbf{M}$ is lowered to $\begin{pmatrix} J_x & \Gamma_{xy} & \Gamma_{xz} \\ \Gamma_{xy} & J_y & \Gamma_{yz} \\ \Gamma_{xz} & \Gamma_{yz} & J_z \end{pmatrix}$, where the Kitaev coupling is now defined by $K = J_z - (J_x + J_y)/2$. The sense of the exchange anisotropy is determined by the sum of the off-diagonal couplings $\Sigma\Gamma = \Gamma_{xy} + \Gamma_{xz} + \Gamma_{yz}$ with positive (negative) values indicative of easy-plane (easy-axis). In the bulk, the large out-of-plane critical field stems primarily from the large off-diagonal $\Gamma > 0$ term, which is the main competitor to the Kitaev interaction.

To correlate the distortions with microscopic interactions, we have performed *ab initio* calculations of the spin Hamiltonian for a range of distortions and evaluated the classical ground state magnetic order, schematics of which are shown in the upper part of Fig. 5c (see Methods and Supplementary Section 5). The results are shown in the lower panels of Fig. 5c as two sets of false-color plots for $\Sigma\Gamma$ as a function of the Cl distortions. The left (right) panel is calculated without (with) Ru buckling. The plots also map out a phase diagram for the magnetic ordering. Regions where classical stripy (Str), ZZ, and ferromagnetic (FM) phases compete have been theorized to realize a QSL state in the bulk[10,20]. The position of bulk α-RuCl$_3$ is marked by the black circle in the left panel[35,36], while the dashed rectangle in the right panel outlines our 1L α-RuCl$_3$ within the error limits of electron diffraction. We have also used density functional theory to calculate the relaxed structure of the freestanding monolayer (see Methods and Supplementary Section 5), which appears near that of the experimental bulk structure and does not exhibit Ru buckling (see red circle, left panel). The microscopic origin of the observed buckling is therefore left as an open question.

Hashed areas in the phase diagram on the right of Fig. 5c mark regions of within the ZZ state within the dashed rectangle where the magnetic anisotropy has flipped to out-of-plane, which all lie on the border to FM order. To narrow the 1L phase boundary further, we have performed magnetic circular dichroism measurements on 1L α-RuCl$_3$ to measure the out-of-plane magnetization and the results are inconsistent with a ferromagnetic (FM) phase with easy axis anisotropy (see Methods and Supplementary Section 6), indicating that our monolayers most likely retain the ZZ configuration and possess a value of $\Sigma\Gamma$ that is small and negative (hence reside in the hashed region). The various exchange terms estimated for this region as well as for the bulk structure are summarized below in Table 1. We thus see that the anisotropy reversal in monolayer samples is largely driven by the in-plane Cl distortion, which suppresses and reverses the off-diagonal exchange. Similar analysis of the *g*-factor supports this conclusion (see Supplementary Section 5). 1L α-RuCl$_3$ appears to be near a transition to out-of-plane FM ordering as a result. Due to out-of-plane Cl compression relative to the bulk structure, 1L α-RuCl$_3$ also lies closer to the region where Str, ZZ, and out-of-plane FM phases compete.

Table 1: Summary of the estimated exchange couplings (meV) for 1L and bulk α-RuCl$_3$.

|  | J | K | Γ | Γ′ | | |
|---|---|---|---|---|---|---|
| No buckling (bulk α-RuCl$_3$) | -3.3 | -6.4 | 3.6 | -0.7 | | |
|  | $J_x$ | $J_y$ | $J_z$ | $Γ_{xy}$ | $Γ_{yz}$ | $Γ_{xz}$ |
| Ru buckling (1L α-RuCl$_3$) | -3.1 | -4 | -11.8 | 2.3 | -4.5 | 1.4 |

In conclusion, our tunneling measurements on 2D α-RuCl$_3$ reveal the presence of single- and two-magnons down to the monolayer limit and a magnon continuum in 3L. The evolution of magnons with magnetic field indicates a clear change in the magnetic anisotropy from easy-plane to easy-axis in monolayer form that is supported by magnetotransport measurements in a gated lateral geometry. 3D electron diffraction shows that 1L α-RuCl$_3$ possesses several structural distortions, among which an in-plane Cl distortion predominantly drives the anisotropy reversal. This is supported by *ab initio* calculations, which are also used to extract a microscopic spin Hamiltonian and distortion-dependent magnetic phase diagram. Relative to the bulk, the ground state of monolayer α-RuCl$_3$ lies in closer proximity to the intersection of several competing spin orders, from which novel Kitaev physics may potentially emerge. Furthermore, while a field-induced QSL for in-plane fields in bulk α-RuCl$_3$ remains a subject of intense debate, a variety of theoretical works have predicted QSL phases for out-of-plane fields that have hitherto been inaccessible due to the large easy-plane anisotropy[6,20–23]. Such states may now be potentially realized for monolayer samples. Our results demonstrate the importance of dimensionality in tuning magnetism in strongly correlated spin systems and pave the way for versatile experimental knobs used for 2D materials (electric field, doping, strain etc.) to further modify the magnetic order in atomically thin α-RuCl$_3$.

# Methods

**Crystal Synthesis**

α-RuCl$_3$: α-RuCl$_3$ single crystals were grown using the chemical vapor transport method. First, the commercial RuCl$_3$ powders was dehydrated at 473K for 12 hours in a dynamic vacuum. Then, dry RuCl$_3$ powder was put into a silica tube with length of 20cm. The tube was evacuated down to $10^{-2}$Pa and sealed under vacuum. The source zone was raised to 923K, and the growth zone was raised to 823K. The growth period was about seven days, and then the furnace was cooled naturally. The shiny black plate-like single crystals of α-RuCl$_3$ can be obtained.

1T′-MoTe$_2$: 1T′-MoTe$_2$ single crystals were grown by the flux method using Te as a solvent. Mo [Alfa Aesar, 99.9%], Te [Alfa Aesar, 99.99%] powders were ground and placed into alumina crucibles in a ratio of 1:25 and sealed in a quartz ampoule. After the quartz ampoule was heated to 1050°C and held for 2 days, the ampoule was slowly cooled to 900 °C over 120 hours and centrifuged. Shiny and plate-like crystals with lateral dimensions up to several millimeters were obtained.

**Device fabrication (IETS, lateral measurement, electron diffraction)**

α-RuCl$_3$, graphite/graphene (HQ Graphene), h-BN (HQ Graphene), and 1T′-MoTe$_2$ were exfoliated on polydimethylsiloxane-based gel (Gel-Pak) within a nitrogen-filled glovebox ($P_{O_2}, P_{H_2O} < 0.1$ppm). Contact electrodes (17nm Au/3nm Ti) and wirebonding pads (40nm Au/ 5nm Ti) are prepatterened by conventional photolithography and e-beam deposition. Device heterostructures for IETS (hBN/MoTe$_2$/α-RuCl$_3$/MoTe$_2$/hBN), gated lateral transport (Gr/hBN/Gr/1L α-RuCl$_3$/Gr/hBN/Gr), electron diffraction (1L Gr/1L α-RuCl$_3$/1L Gr) samples were sequentially stacked by polycarbonate films at 90°C in the glovebox. To prevent electrical breakdown of the atomically thin α-RuCl$_3$, the current should be minimized in IETS measurements, and so the junction area is kept small (around 0.3μm$^2$, 1.5μm$^2$, and 5μm$^2$ for 1L, 2L, and 3L α-RuCl$_3$, respectively).

**Magnetotransport measurements**

Magnetotransport measurements were mostly performed in a superconducting magnet He-4 cryostat (base temperature 1.4K, magnetic field limit 14T). A superconducting magnet He-3 cryostat (base temperature 0.3K, magnetic field limit 12T) was used for an IETS device with Gr contacts. Both setups have a single-axis rotator for the sample stage. DC measurements were performed using a Keithley 2450 source measure unit. DC + AC measurements were performed using a combination of a Keithley 2450 source measure unit and SRS 830/860 lock-in amplifiers.

**3D electron diffraction measurements**

Acquiring 3D electron diffraction patterns was accomplished by tilting the specimen over a range of angles relative to the incident beam to provide slices through the reciprocal structure. Selected area electron diffraction (SAED) patterns were acquired on the TFS Talos F200X G2 operating at 80keV with TEM holder tilting the sample from +35° to -35° in 1° increment. An accelerating voltage of 80keV was chosen to minimize beam induced damage to the 2D material. A 0.75μm SAED aperture was centered over the same sample region throughout the tilt series acquisition. Each SAED in the tilt series is first background subtracted and aligned to a common center. Diffraction spots pertaining to α-RuCl$_3$ at every specimen tilt were characterized by fitting a four-parameter two-dimensional Gaussian to a windowed region about each peak. The integrated diffraction peak intensity was then calculated and plotted against $k_z$ for curve fitting with the kinematic model.

**Raman Spectroscopy**

Raman spectroscopy was carried out at room temperature using a 532nm excitation laser in backscattering geometry with a beam spot size of ~1μm. The laser power was kept at ~0.1μW, to minimize the local heating effect. The scattered light was dispersed by a Horiba LabRAM HR Evolution Raman Microscope system and detected by a thermoelectric cooled CCD camera. The hBN-encapsulated α-RuCl$_3$ flakes were mounted on a rotatable stage and measured at every 10°.

**Magnetic Circular Dichroism**

The magnetization of hBN-encapsulated 1L α-RuCl$_3$ flakes was characterized by magnetic circular dichroism microscopy in a superconducting magnet He4 cryostat (AttoDry1000) with out-of-plane magnetic field. A diode laser at 410nm with an optical power of ~10μW was focused onto a submicron spot on the flakes using an objective with numerical aperture 0.8. The optical excitation was modulated by a photoelastic modulator at ~50kHz for left and right circular polarization. The laser reflected from α-RuCl$_3$ was collected by the same objective and then detected by a photodiode. The MCD signal is defined as the ratio of the modulated signal (measured by a lock-in amplifier) to the total reflected light power (measured by a DC voltmeter).

**Ab-initio calculations**

**1.Magnetic Couplings**

In order to estimate the magnetic couplings, we employed the exact diagonalization method outlined in ref[11,64]. Hopping integrals, crystal field tensors, and spin-orbit coupling in the basis of the five Ru *4d* orbitals were first computed for each structure using the density functional theory package FPLO[65] at the fully relativistic GGA (PBE) level. For structures without Ru buckling, we employed an idealized monolayer structure with *P-312/m* symmetry and a large vacuum gap between monolayers. The in-plane lattice constant was set to 5.979Å, which is consistent with the results of electron diffraction. To simulate the Ru buckling, we repeated the calculations with $\Delta\zeta_{Ru} = 0.3$Å, representing the best fit from electron diffraction (formally lowering the symmetry to *P-3*). For each structure, the computed one-particle terms were used to define a two-site model with Hamiltonian given by $H = H_{1-p} + H_U$ where the Coulomb interactions $H_U = \sum_{\sigma,\sigma'} \sum_{\alpha\beta\gamma\delta} U_{\alpha\beta\gamma\delta} c^\dagger_{i,\alpha,\sigma} c^\dagger_{i,\beta,\sigma'} c_{i,\gamma,\sigma'} c_{i,\delta,\sigma}$ were defined in the spherically symmetric approximation defined in terms of the Slater parameters $F_0$, $F_2$, and $F_4$[66]. For this purpose, we use $U_{t2g} = 2.58$, $J_{t2g} = 0.29$eV following the ref[67], and approximate $F_4/F_2 = 5/8$[68]. This corresponds to $F_0 = 2.15$eV, $F_2 = 3.24$eV, and $F_4 = 2.02$eV. After exactly diagonalizing the two-site model, we extract the magnetic couplings by projecting onto pure $j_{1/2}$ doublets of the ideal $d^5$ ground state.

**2. g-Tensors**

In order to estimate the magnetic *g*-tensors, we employed the method outlined in ref[69]. From the structures employed in the calculation of the magnetic interactions, we extracted the coordinates of a single $[RuCl_6]^{3-}$ octahedron. For each, we computed the *g*-tensors using ORCA[70] at the def2-SVP/PBE0/CAS-SCF(3,5) level. This approach has proved reliable in previous studies of $RuCl_3$ and other materials, and is consistent with expected trends.

**3. DFT Structural Relaxation**

Our structural relaxation calculation of monolayer α-$RuCl_3$ is based on spin-polarized density functional theory (DFT) as implemented in VASP code[71] with a generalized gradient approximated (GGA) exchange-correlation functional. The interaction between ion cores and valence electrons is described by pseudopotential of projector augmented wave (PAW) type. A correction due to van der Waals forces are included through the DFT-D2 scheme of Grimme[72]. A plane-wave cutoff of 600eV is used for the 2x2 supercell in the slab geometry with 3x3x1 k-point sampling. The in-plane lattice parameters (a = b = 12.00Å for 2x2 supercell) are chosen based on the electron diffraction results. A minimum distance of 9 Å is kept between two periodic images along c-direction.

# Acknowledgements

A.W.T. acknowledges support from the US Army Research Office (W911NF-21-2-0136), Ontario Early Researcher Award (ER17-13-199), and the National Science and Engineering Research Council of Canada (NSERC) (RGPIN-2017-03815). This research was undertaken thanks in part to funding from the Canada First Research Excellence Fund. R.H., S.S.H., Y.M.G. acknowledge support from the Keck Foundation. S.M.W. was supported by a pilot grant from the Center for Functional Materials, and performed computations on the Wake Forest University DEAC Cluster, a centrally managed resource with support provided in part by Wake Forest University. D.K., S.B. and R.V. acknowledge support by the Deutsche Forschungsgemeinschaft (DFG) through grants VA 1171/15-1 and TRR 288-4 22213477 (Project A05). G. Y. and R. H. are supported by NSF CAREER Grant No. DMR-1760668 and NSF Grant No. DMR-2104036. H.C.L. acknowledges support by the National Key R&D Program of China (Grant No. 2018YFE0202600), the Beijing Natural Science Foundation (Grant No. Z200005), and the Fundamental Research Funds for the Central Universities, and the Research Funds of Renmin University of China (18XNLG14, and 19XNLG17).

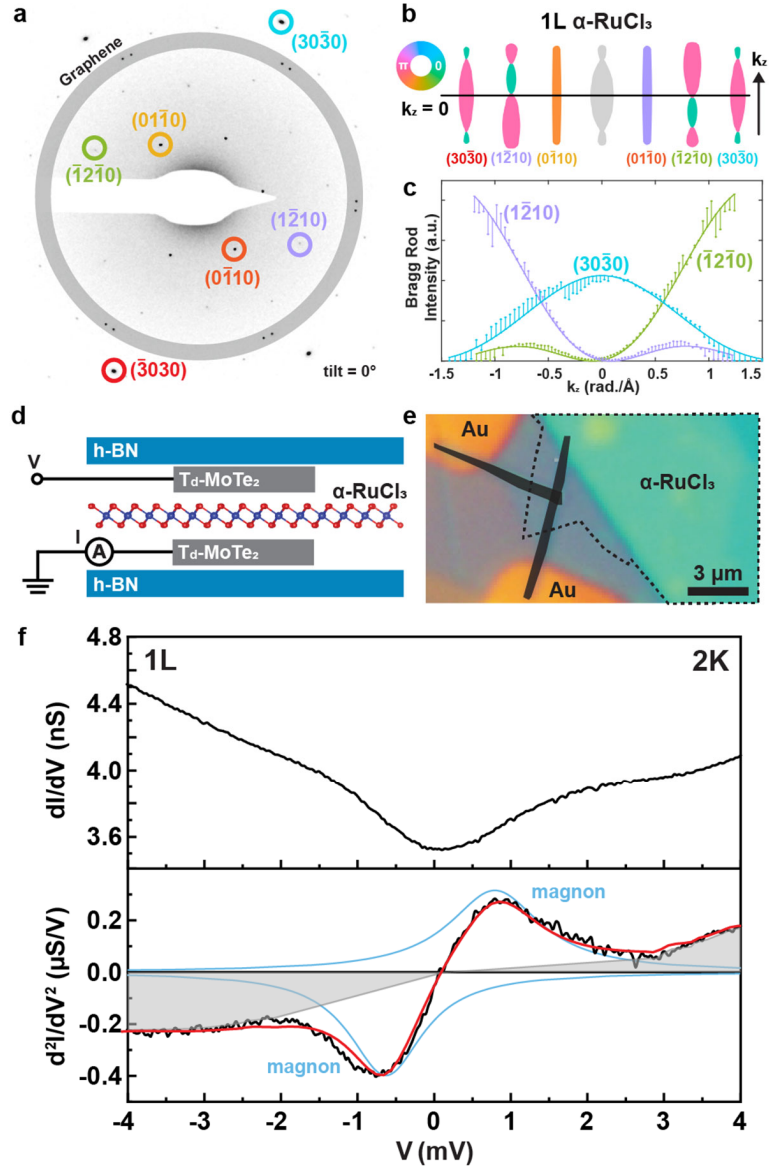

**Figure 1 | 3D electron diffraction and demonstration of IETS measurements on 1L α-RuCl$_3$.** (a): Electron diffraction pattern for graphene-encapsulated 1L α-RuCl$_3$ at 0° tilt. Bragg peaks for graphene layers are marked by a thick gray circle. Several α-RuCl$_3$ Bragg peaks selected for analysis are circled. (b): Schematic of calculated out-of-plane momentum ($k_z$) dependence for the various Bragg rods of 1L α-RuCl$_3$ chosen in (a). The thickness and color indicate the complex magnitude and phase of the structure factor, respectively. (c): Experimental Bragg intensities (scatter points) for (1$\bar{2}$10), ($\bar{1}$2$\bar{1}$0), and (30$\bar{3}$0) peaks, plotted as function of $k_z$, show great agreement with fitted kinematic model (lines) of 1L α-RuCl$_3$. (d): Sideview schematic of an IETS device with vertical T$_d$-MoTe$_2$ contacts to few-layer α-RuCl$_3$. (e): Colorized optical image of a 1L α-RuCl$_3$ device. Black shaded areas represent T$_d$-MoTe$_2$ and dashed lines outline α-RuCl$_3$ flake. (f): Representative IETS results for 1L α-RuCl$_3$ taken at 2K. Upper panel: AC tunneling conductance dI/dV as a function of applied DC voltage showing subtle steps due to magnon excitations at both positive and negative voltage. Lower panel: Numerical derivative (black trace) of experimental dI/dV curve, dI$^2$/d$^2$V, together with results from fitting (gray: background; blue lines: Lorentzian fits to magnon peaks; red: overall fit).

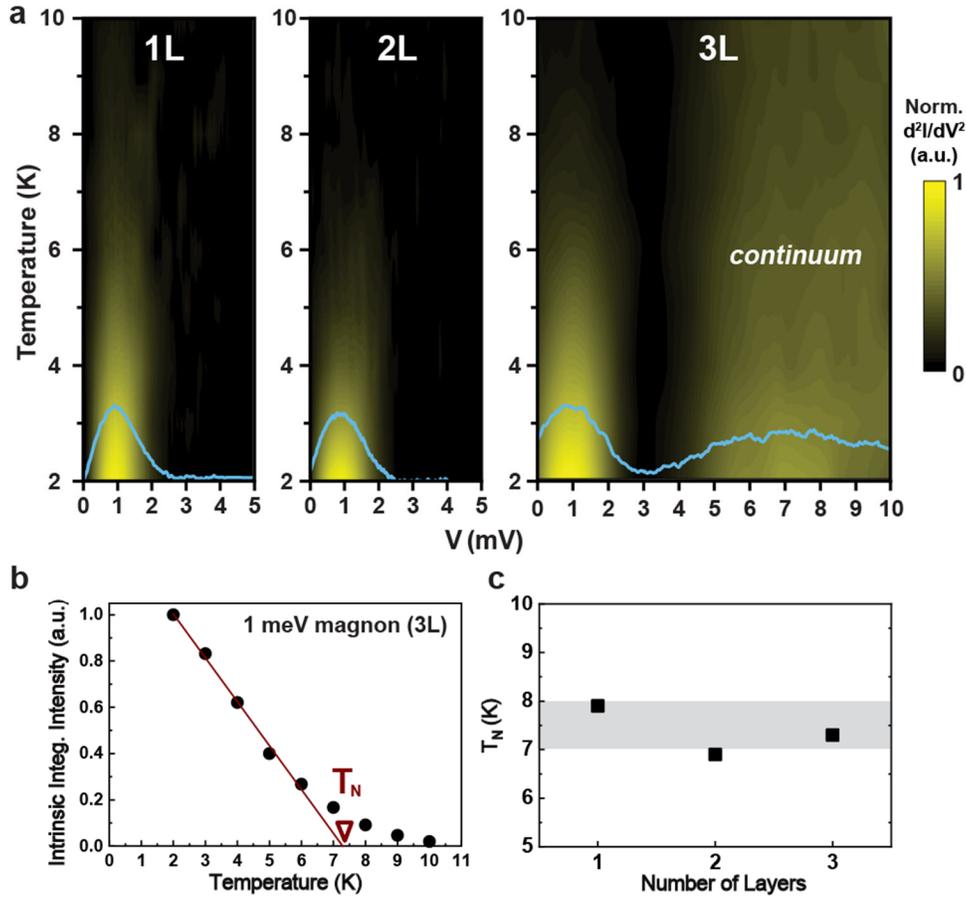

**Figure 2 | Temperature-dependent IETS on few-layer α-RuCl$_3$.** (a): False-color plot of normalized and background-subtracted $dI^2/d^2V$ spectra for 1L, 2L, and 3L α-RuCl$_3$ from 2K to 10K at positive DC bias. The trace at 2K for each thickness is overlaid in blue. A broad excitation between ~5meV to ~10meV is observed for 3L and attributed to the magnon continuum. (b): Intrinsic integrated intensity for ~1meV magnon of 3L α-RuCl$_3$ at each temperature. The data points from 2K to 6K are utilized for linear fitting (red line), whose x-intercept yields $T_N$ ~ 7K. (c): Thickness-dependent $T_N$ extracted using the same procedure in (b) for 1L, 2L, and 3L α-RuCl$_3$, all of which fall in the range of 7K-8K (gray band), which corresponds to the range reported for high-quality bulk crystals.

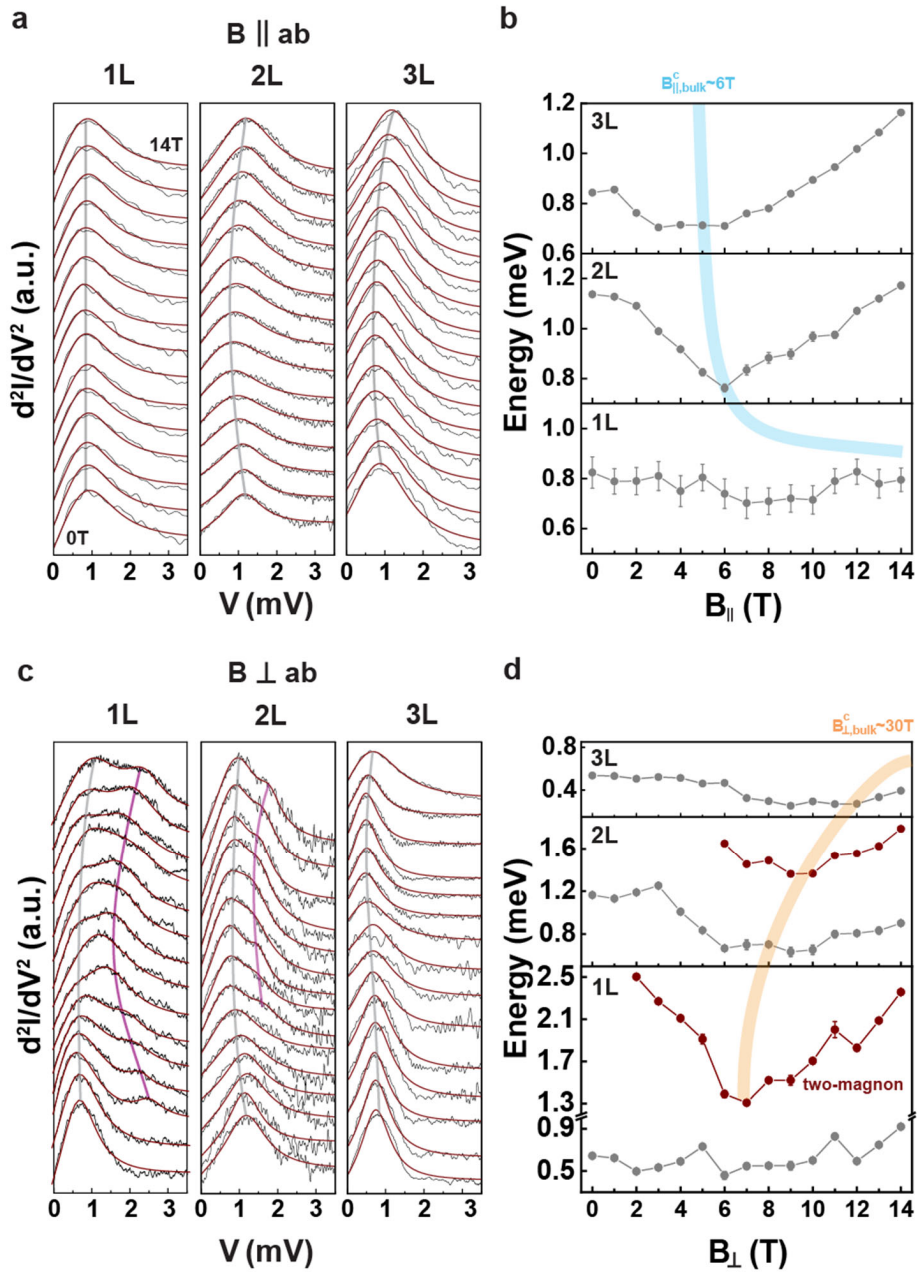

**Figure 3 | Magnetic-field-dependent IETS on few-layer α-RuCl₃.** 1L, 2L, and 3L α-RuCl₃ spectra (black lines) with changing $B_\parallel$ (a) and $B_\perp$ (c) from 0T to 14T in 1T increments and offset for clarity. Lorentzian fitting and background subtraction is performed for each $dI^2/d^2V$ trace (red lines). The overall trend of the magnon evolution is marked by thin gray and purple lines. The extracted peak positions for each thickness are plotted as function of $B_\parallel$ (b) and $B_\perp$ (d). Gray points are results for low-energy magnon and red points are for two-magnon scattering). Thick blue and orange lines capture trends in changing critical field with thickness for $B_\parallel$ and $B_\perp$, respectively.

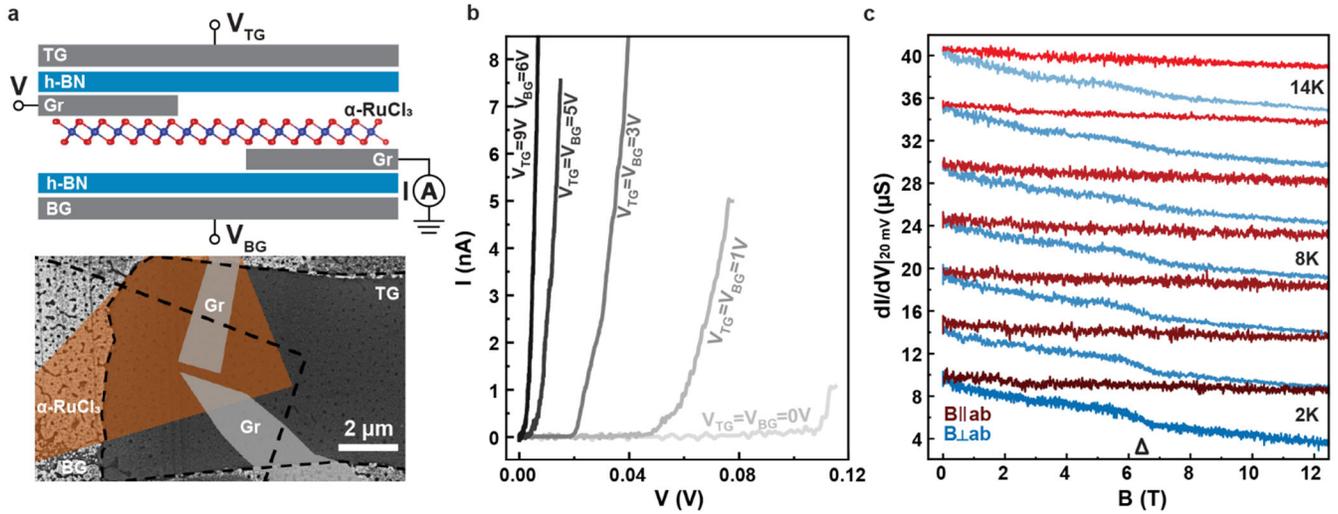

**Figure 4 | Lateral magnetotransport measurement on 1L α-RuCl$_3$ with dual gates.** (a): Sideview schematic (upper) and colorized scanning electron microscopy image (lower) of device with channel length of ~300nm. (b): DC current-voltage characteristics curves at 2K taken with various gate voltages. (c): AC conductance at 20mV DC bias, 9V top gate, and 6V back gate with changing with $B_{||}$ (red) and $B_\perp$ (blue) from 2K to 14K in 2K increments. The magnetoconductance has larger change overall with $B_\perp$ and a marked kink at $B_\perp \sim 6.5$T that gradually disappears above $T_N \sim 8$K.

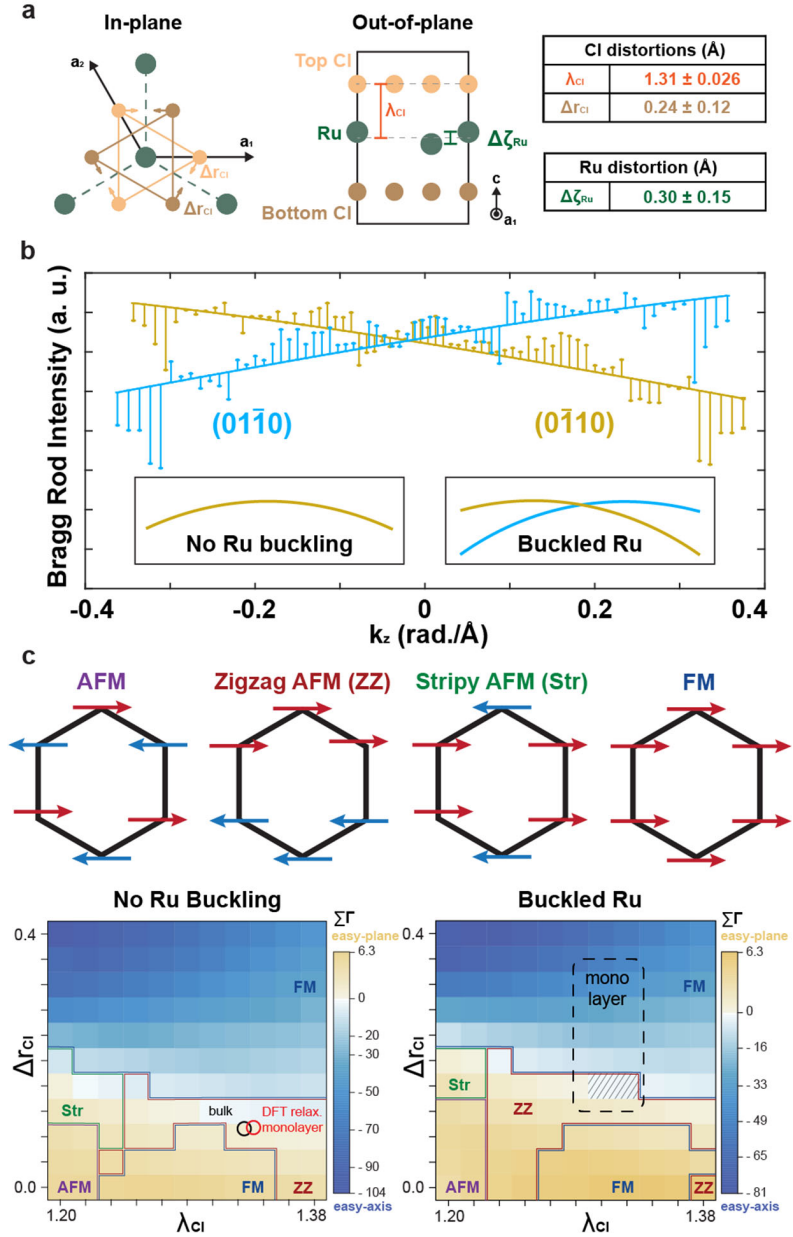

**Figure 5 | Three primary distortions of 1L α-RuCl$_3$ and magnetic phase diagram.** (a): Schematic illustration of three distortions (left) and summary of the values determined by 3D electron diffraction measurements (right). (b): Asymmetric $(01\bar{1}0)$ and $(0\bar{1}10)$ Bragg rod intensities vs $k_z$ indicate out-of-plane Ru buckling. Inset: schematic for the two Bragg rods with and without Ru buckling. (c): Upper panel: magnetic phase diagram determined by *ab initio* calculations of ΣΓ as a function of the Cl distortions without Ru buckling ($\Delta\zeta_{Ru} = 0$ Å, left) and with Ru buckling ($\Delta\zeta_{Ru} = 0.3$ Å, right). Yellow (blue) regions correspond to easy-plane (easy-axis) magnetic anisotropy. Positions for the experimental bulk structure and DFT-relaxed monolayer are circled in the case of no buckling. The dashed rectangle in the right panel outlines our 1L α-RuCl$_3$ within the error limits of electron diffraction. Hashed areas mark regions within the rectangle where the ZZ magnetic anisotropy has flipped to out-of-plane (easy-axis). Lower panel: Schematics of various classical magnetic orders in phase diagram.

23